\documentclass[10pt,journal,compsoc]{IEEEtran}
\usepackage{graphicx} 
\usepackage{booktabs}
\usepackage{tabularx}     
\usepackage{adjustbox} 
\usepackage{pgfplots}
\usepackage{subcaption}  
\usetikzlibrary{trees, positioning, shapes, arrows.meta}
\usepackage{xcolor}   
\pgfplotsset{compat=1.18}
\usepackage{tikz}
\usepackage{amsmath}
\usepackage{booktabs,caption}
\usepackage[flushleft]{threeparttable}
\usepackage{siunitx}
\usetikzlibrary{positioning, shapes.multipart, arrows.meta}
\usepackage{catchfilebetweentags}
\usepackage[colorlinks=true,linkcolor=blue,citecolor=green]{hyperref}
\usepackage{epigraph}

\begin{document}

\author{Temitayo Adefemi}

\IEEEtitleabstractindextext{%
\begin{abstract}
Ever since Claude Shannon used entropy for his "Mathematical Theory of Communication", entropy has become a buzzword in research circles with scientists applying entropy to describe any phenomena that are reminiscent of disorder. In this paper, we used entropy to describe the incompatibility between components in the computer, which can cause noise and disorder within the parallel cluster. We develop a mathematical theory, primarily based on graph theory and logarithms, to quantify the entropy of a parallel cluster by accounting for the entropy of each system within the cluster. We proceed using this model to calculate the entropy of the Top 10 supercomputers in the Top500 list. Our entropy framework reveals a statistically significant negative correlation between system entropy and computational performance across the world's fastest supercomputers. Most notably, the LINPACK benchmark demonstrates a strong negative correlation (r = -0.7832, p = 0.0077) with our entropy measure, indicating that systems with lower entropy consistently achieve higher computational efficiency, this Relationship is further supported by moderate correlations with MLPerf mixed-precision benchmarks (r = -0.6234) and HPCC composite scores (r = -0.5890), suggesting the framework's applicability extends beyond traditional dense linear algebra workloads.
\end{abstract}

\begin{IEEEkeywords}
Entropy, Parallel Computing, Distributed Computing, Information Theory
\end{IEEEkeywords}}

\title{The Entropy of Parallel Systems}
\maketitle

\epigraph{"You should call it entropy, for two reasons. In the first place, your uncertainty function has been used in statistical mechanics under that name. In the second place, and more importantly, no one knows what entropy really is, so in a debate you will always have the advantage."}{\textit{John Von Neumann quote to Claude Shannon}}

\vspace{0.9cm}
\IEEEdisplaynontitleabstractindextext

%
\IEEEpeerreviewmaketitle

\ifCLASSOPTIONcompsoc
\IEEEraisesectionheading{\section{Introduction}\label{sec:introduction}}
\else
\section{Introduction}
\label{sec:introduction}
\fi

\IEEEPARstart{O}{ne} might ask if there is disorder in parallel systems when solving a parallel problem, and how can this disorder be understood, conceptualised, and calculated? A basis would be that the more computers in a cluster, the more disorder there should be when solving a parallel problem, but that is just a baseline, and the entropy of parallel systems should be more sophisticated and encompassing; the entropy of a parallel computing cluster should give insight into whether we should use that cluster to solve the problem or even build the parallel cluster in the first place. It should provide a foundation that allows computer scientists to access parallel systems holistically and make informed decisions based on the analysis. Entropy is a concept that has been used extensively in several fields, including quantum physics and information theory \cite{shannon}, \cite{neumann}. The concept was made well known by Claude Shannon with his use of entropy to formulate his "Mathematical Theory of Communication," in which he used entropy to measure the perplexity in information.

\vspace{0.3cm}
Concepts used to control the entropy of distributed systems are the same tools and patterns used to guarantee synchronisation. However, not all parallel systems are distributed; even if they can be, there is currently no formal definition of entropy in computer science for both distributed and parallel systems.\cite{vishesh}.

\vspace{0.3cm}

There should be several points that should be considered when developing a theory to conceptualize the entropy of parallel systems, and these include:
\begin{itemize}
    \item How compatible are the computers included in the cluster? And to what degree?
    \item How compatible are the components of a computer, CPU, Memory, Interconnect, GPU, etc., with each other?
    \item How many computers are included?
\end{itemize}

These factors are the basic intuitions that guide the mathematical theory used in this paper to aid in calculating the entropy of parallel systems.

\vspace{0.3cm}
\textbf{Goal.} Create a mathematical theory that gives computer scientists and high performance computing specialists tools that can be used to calculate the entropy of parallel systems when solving a parallel problem, to give insight into whether the problem is worth solving on the parallel cluster and whether alternatives should be considered.

\vspace{0.3cm}
\textbf{Approach.} We harness the power of \textbf{graph theory}, and \textbf{logarithms} to develop a mathematical theory to calculate the \textbf{entropy} ($H$) of a parallel system based on the entropy of each computer in the cluster.

\vspace{0.3cm}

We model each computer as a \textbf{bidirectional graph} $G=(V, E)$ of its components, where $V$ is the set of nodes (components) and $E$ is the set of edges representing the communication between nodes. We give each node $v \in V$ a base value of $10$, then process every bidirectional edge based on a compatibility matrix, and then we take the most significant value from the set of values in the graph, which we will denote as $v_{\text{max}}$.

\vspace{0.3cm}

We use the most significant value of each computer in a cluster, denoted as $S = \{v_{\text{max}_1}, v_{\text{max}_2}, \dots, v_{\text{max}_n}\}$, where $n$ is the number of computers. We then sum the values from the \textbf{logarithm penalty function} to return the value of the entropy of the parallel cluster, $H_{\text{cluster}}$. The formula gives this:

\[
S_{\text{parallel}} = \sum_{i=1}^{n} P(S_i) = \sum_{i=1}^{n} 3 \log(1 + S_i)
\]

\vspace{0.3cm}

We use this mathematical theory to calculate the entropy of the Top 10 supercomputers on the Top500 supercomputers list, showing how this tool can be used to gauge the performance of the benchmarks run on these systems  We report performance and show correlation data from the LINPACK benchmark, which measures computing power by assessing how quickly a system can solve a dense system of linear equations—a prevalent task in scientific computing and engineering~\cite{dongarra}.

\vspace{0.3cm}

There are other benchmarks which report performance to complement the LINPACK, including: 
\begin{itemize}
\item STREAM: a synthetic benchmark program that measures sustainable memory bandwidth (in GB/s) and the corresponding computation rate for simple vector kernels~\cite{john}
\item High-Performance Conjugate Gradient (HPCG): which examines data access patterns of real-world applications such as sparse matrix calculations, thus testing/stressing memory subsystems~\cite{dongarra}
\item Several other benchmarks, including HPCC~\cite{luszczek}, MLPerf HPC~\cite{mattson}, Graph500~\cite{ueno} and HPCAI \cite{jiang}
\end{itemize}

\section{Related Work}
Insufficient studies have been conducted to quantify the entropy or disorder of parallel systems. In this section, we will focus on a sparse number of research that contribute to and surround this work, starting from the concept of entropy, followed by how researchers have applied this abstract idea to understanding parallel systems.

\vspace{0.3cm}

"The Mathematical Theory of Communication" by Claude Shannon was the piece of research that made entropy a household name in research circles \cite{shannon}. Entropy had been previously utilized in statistical dynamics to measure uncertainty, disorder, or mixedness in the phrase of Gibbs, which remains about a system after its observable macroscopic properties, such as temperature, pressure, and volume, have been taken into account. After Claude Shannon released his theory, researchers from numerous fields started applying the idea to a wide range of fields, including thermodynamics and quantum mechanics \cite{shannon} \cite{popovic} \cite{neumann}.

\vspace{0.3cm}

Claude Shannon used entropy to measure the uncertainty or randomness of a system. It quantifies the average amount of information needed to represent an event or outcome from a probability distribution. Higher entropy signifies greater unpredictability and more information content, while zero entropy indicates an entirely predictable event with no uncertainty. Information entropy provides a theoretical limit on the efficiency of lossless data compression and transmission \cite{shannon} \cite{willems}.

\vspace{0.3cm}

In thermodynamics, entropy is a quantitative measure of the disorder, randomness, or "spread-out-ness" of a system's energy and matter. It is a state function that also indicates the amount of energy unavailable to perform valuable work. A system with higher entropy is more disordered, which means its energy is more dispersed, and there are more possible microscopic arrangements (microstates) that describe that state \cite{popovic}.

\vspace{0.3cm}

We also have quantum entropy, which is also called the von Neumann entropy. It quantifies the amount of classical uncertainty about the quantum state of a system. Suppose a system A is in a pure state, meaning that it has a well-defined value for some property (e.g., it is a horizontally polarized photon) \cite{neumann}.

\vspace{0.3cm}
The only relevant research undertaken to study entropy in parallel systems is a theory that was developed for visualizing the parallel execution by the entropy of the phase space induced by its traces. The metric also showed that it is possible to both theoretically and practically find program issues, such as starvation due to data in one of its threads \cite{gomez}.

\section{Mathematical Theory for the Entropy of Parallelization}

We have devised mathematical formulations that would be used to calculate the entropy of parallel systems based on the incompatibility of interconnected components. Let $S_{\text{total}}$ be the entropy of the parallel system as a whole, and $S_i$ be the entropy of the $i$-th machine in the cluster. The total entropy is a function of the individual machine entropies, $S_{\text{total}} = f(S_1, S_2, \ldots, S_n)$. We first focus on calculating the entropy of a single machine, which we will denote as $S_{\text{computer}}$.

\subsection{Entropy of a Single Machine in the Cluster}
To calculate $S_{\text{computer}}$, the system's components are modeled as a bidirectional graph $G = (V, E)$. The set of vertices, $V$, represents the hardware components, while the set of edges, $E$, represents the communication between them. The vertex set is defined as $V = \{v_{\text{CPU}}, v_{\text{GPU}}, v_{\text{Cache}}, v_{\text{Memory}}, \}$. We put a bidirectional edge $(u, v) \in E$ between every component $u$ in the architecture and $v$. Manufacturers of high-performance computing components also frequently develop their own compiler software, and compatibility can be critical and deterministic for the performance of parallel computing software. Integrating the compiler into the system would be beneficial and provide a more accurate representation of the computer's entropy value. Still, for the first iteration of this model, we would solely focus on the Von Neumann architecture components of the computer.

\vspace{0.3cm}

We also exclude microarchitectural features such as NUMA, memory hierarchy depth, network topology, PCIe/NVLink, and fabric generation from edges and manufacturers—many critical factors (OS, drivers, firmware, MPI stack, PCIe/NVLink versions). The reasoning behind this decision is that, even though these hardware and software components can have a significant impact on performance, modelling the entire computer system of interconnected parts would be counterintuitive. However, further research in this area could focus on considering additional factors.

\vspace{0.3cm}

Each vertex $v \in V$ is assigned a constant base value, $B(v)$. For this model, we set $\forall v \in V, B(v) = 10$. There are no weights on the edges. This base value is chosen to create a great starting point. By setting a relatively large base value like 10, the interaction formula ensures that even a minor incompatibility (a compatibility score closer to 0) produces a significantly high interaction value. This approach realistically models how a single inefficient component can bottleneck an entire system, thereby validating the choice of using the maximum interaction value to represent a computer.

\vspace{0.3cm}

Additionally, each component $v$ has a manufacturer, denoted by $M(v)$. To move beyond a binary compatible/incompatible model, we introduce a graded approach using a \textbf{Compatibility Matrix, $C$}. This matrix provides a compatibility score, $C(M_1, M_2)$, for any pair of manufacturers, with values ranging from 0 (incompatible) to 1 (perfectly compatible).

\vspace{0.3cm}

For any given edge $(u, v) \in E$, we calculate an interaction value, $I(u, v)$, using the following formula:
\[
I(u, v) = \frac{B(u) \times B(v)}{C(M(u), M(v)) + \epsilon}
\]
Where $C(M(u), M(v))$ is the compatibility score sourced from the matrix, and $\epsilon$ is a small constant (e.g., $10^{-9}$) to prevent division by zero. This formulation creates a spectrum of interaction values. For instance, with $B(v) = 10$:
\begin{itemize}
    \item If components are perfectly compatible ($C=1.0$), $I(u,v) = \frac{10 \times 10}{1} = 100$.
    \item If components have high but not perfect compatibility (e.g., $C=0.9$), $I(u,v) = \frac{100}{0.9} \approx 111$.
    \item If components are severely incompatible (e.g., $C=0.01$), $I(u,v) = \frac{100}{0.01} = 10000$.
\end{itemize}

This approach realistically models that minor incompatibilities create small amounts of disorder, while major ones create significant disorder. The values for the matrix $C$ can be determined through empirical benchmarks or expert heuristics. In the initial iterations of this model, we employed a binary classification to determine whether a component is compatible with another element. However, we decided to use a compatibility matrix because compatibility is far more nuanced and multifaceted, requiring a more sophisticated methodology to calculate.

\vspace{0.3cm}

We process every edge in the graph and calculate its corresponding interaction value. The set of all calculated interaction values is $\mathcal{I} = \{I(e) \mid e \in E\}$.

\vspace{0.3cm}

The entropy of the singular computer, $S_{\text{computer}}$, is then defined as the maximum value from this set:
\[
S_{\text{computer}} = \max(\mathcal{I}) = \max_{(u,v) \in E} \{I(u, v)\}
\]

\vspace{0.3cm}
The rationale for choosing the maximum value is to quantify the highest potential disorder within the system, analogous to how the complexity of an algorithm is determined by its worst-case analysis. Taking the average value of all the nodes is impractical for this analysis, as when a single component bottlenecks a computer, the effect propagates throughout the system. This single bottleneck dwarfs the efficiency of computation, and evidence of this is shown in the result when applying this model to evaluate supercomputers. Norms, spectral quantities, path-based aggregates, cut/flow metrics, and distributional measures were considered as measures to grade the whole graph network of the parallel system. However, the maximum was utilised as we visualised the computer system as a pipeline. That specific match of incompatibility has the potential to produce enormous noise and disorder in the parallel system that can impact efficiency.

\subsection{Entropy of Parallel Cluster}
To calculate the entropy of the parallel cluster, denoted as $S_{\text{parallel}}$, we first gather the set of entropy values from all $n$ computers within the cluster. Let this set be $\mathcal{S}_{\text{cluster}}$, where each element $S_i$ represents the entropy of the $i$-th computer, calculated as previously described.
\[
\mathcal{S}_{\text{cluster}} = \{S_1, S_2, \ldots, S_n\}
\]
We then apply a logarithmic penalty function, $P(x)$, to each individual entropy value $S_i$ in the set. The penalty function is defined as:
\[
P(x) = 3 \log(1 + x)
\]
This transformation converts the initial set of entropies

\begin{align*}
\mathcal{S}'_{\text{cluster}} &= \{P(S_1), P(S_2), \ldots, P(S_n)\} \\
&= \{3 \log(1 + S_1), 3 \log(1 + S_2), \ldots, 3 \log(1 + S_n)\}
\end{align*}
\vspace{0.3cm}

The final entropy of the parallel cluster, $S_{\text{parallel}}$, is obtained by summing all the values in this new set. This can be expressed using summation notation as:
\[
S_{\text{parallel}} = \sum_{i=1}^{n} P(S_i) = \sum_{i=1}^{n} 3 \log(1 + S_i)
\]

\vspace{0.3cm}
We use logarithm penalties to normalise the numbers derived from the entropy values calculated for each single node in a cluster; the values from incompatibility matches could have grown out of proportion from starting at a base value of 10, and these numbers would need a function to convert them to a lower exponent that makes calculations and interactions with these numbers more usable and grounded, we chose base 3 to penalise higher values while still maintaining the minimalism while not comprising the magtitude of the value derived from the entropy calculation of each node in the cluster.

\section{Entropy of the Top 10 supercomputers}
To ground this mathematical theory in empirical data, we use this model to calculate the entropy of the Top 10 supercomputers in the Top 500 and demonstrate a correlation between the entropy value and the performance reported from the LINPACK, STREAM, MLPerf HPC, HPCC, Graph500, and HPCAI benchmarks. We obtained the specifications from publicly accessible data on these supercomputers in order to determine their entropy level.

\vspace{0.3cm}

It is important to note that the compatibility matrix employed in our entropy calculations was constructed through heuristic analysis rather than empirical derivation, representing an initial framework based on reasonable assumptions about real-world system architectures. There is currently no standard for compatibility when it comes to high-performance computing vendors, as most are competitors with each other, so creating an empirically validated compatibility matrix is unfeasible with current methods. We used an intuitive methodology which involves establishing same-vendor diagonal elements reflecting known integration levels—AMD-AMD (0.95) based on MI300A APU design, Intel-Intel (0.88) accounting for newer GPU integration challenges, Fujitsu-Fujitsu (0.98) representing complete vertical integration of A64FX architecture, and HPE/Cray-HPE/Cray (0.95) reflecting mature optimization—while cross-vendor values incorporated strategic partnerships like AMD-HPE/Cray (0.90) from Frontier/El Capitan collaboration, standard integrations such as Intel-NVIDIA (0.82) representing mature heterogeneous pairing, and legacy system experience like IBM-NVIDIA (0.79) from Summit/Sierra deployments.

\vspace{0.3cm}

This framework could be a starting point for this model as we further investigate how to create a robust and empirically validated methodology for high-performance computing specialists to build on. The heuristic framework assumed perfect compatibility as 1.0 (theoretical maximum), same-vendor systems ranging from 0.85 to 0.98, established partnerships from 0.80 to 0.90, standard interfaces from 0.75 to 0.85, and poor integration from 0.70 to 0.80. A proper empirical derivation would require systematic benchmarking across identical workloads on different vendor combinations to establish $C(M_1, M_2)$= f(benchmark\_performance, theoretical\_peak), utilizing STREAM for memory-CPU compatibility, LINPACK for system-wide integration, specialized GPU benchmarks, and communication benchmarks for interconnect analysis, followed by regression analysis and factor analysis to isolate vendor compatibility effects across multiple system generations. The development and maintenance of an empirically confirmed compatibility matrix is essential for subsequent versions of this theory. It is possible to include new vendors into the framework as they emerge methodically. Supercomputers can be evaluated by those who create them using a unified compatibility matrix. The compatibility matrix utilised for the calculations is shown in \ref{compatibility}:

\begin{table*}[h!]
\caption{Compatibility Matrix $C(M_1, M_2)$ for Component Manufacturers}
\label{compatibility}
\centering
\label{tab:compatibility_matrix}
\begin{tabular}{|c|c|c|c|c|c|c|}
\hline
\textbf{Manufacturer} & \textbf{AMD} & \textbf{Intel} & \textbf{NVIDIA} & \textbf{IBM} & \textbf{Fujitsu} & \textbf{HPE/Cray} \\
\hline
\textbf{AMD} & 0.95 & 0.82 & 0.81 & 0.79 & 0.75 & 0.90 \\
\hline
\textbf{Intel} & 0.82 & 0.88 & 0.82 & 0.78 & 0.74 & 0.85 \\
\hline
\textbf{NVIDIA} & 0.81 & 0.82 & 0.92 & 0.79 & 0.73 & 0.87 \\
\hline
\textbf{IBM} & 0.79 & 0.78 & 0.79 & 0.85 & 0.72 & 0.80 \\
\hline
\textbf{Fujitsu} & 0.75 & 0.74 & 0.73 & 0.72 & 0.98 & 0.76 \\
\hline
\textbf{HPE/Cray} & 0.90 & 0.85 & 0.87 & 0.80 & 0.76 & 0.95 \\
\hline
\end{tabular}
\end{table*}

\begin{figure*}[h!]
    \centering
    \includegraphics[width=1\textwidth]{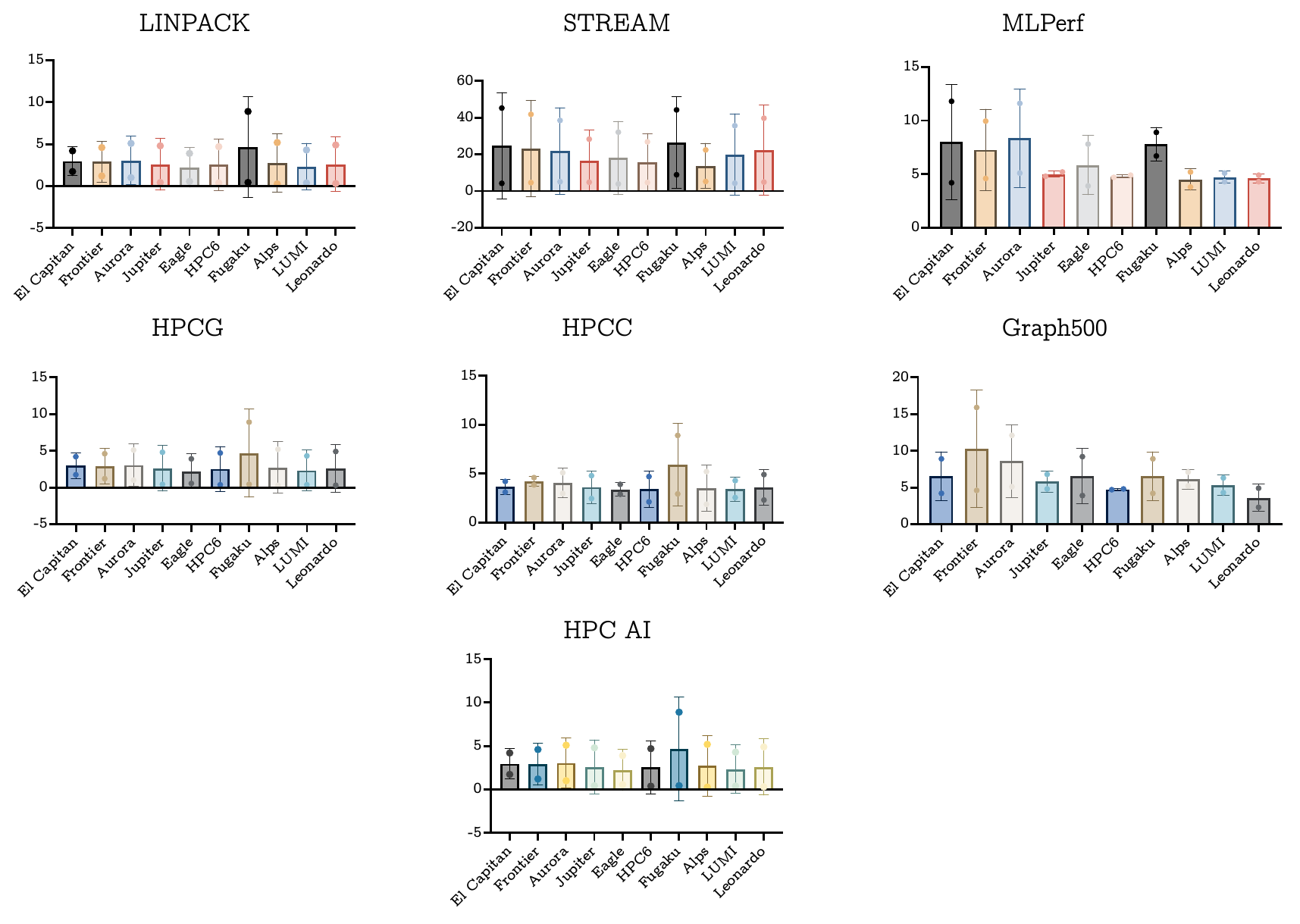}
    \caption{Relationship between the entropy level and performance of selected HPC benchmarks}
    \label{fig:your-label}
\end{figure*}

\begin{table*}[h!]
\centering
\caption{Top 10 Supercomputers: Entropy Values and Benchmark Performance}
\label{tab:entropy_performance}
\begin{tabular}{||c c c c c c c c c||} 
 \hline
 Supercomputer & Entropy Value & LINPACK & STREAM & MLPerf & HPCG & HPCC & Graph500 & HPCAI\\ 
 & & (EFlop/s) & (TB/s) & (EFlop/s) & (PFlop/s) & (GFlop/s) & (GTEPS) & (Score) \\ [0.5ex] 
 \hline\hline
 El Capitan & 4.2 & 1.742 & 45.2 & 11.8 & 2.79 & 3.21 & 8.9 & 15.2 \\ 
 \hline
 Frontier & 4.6 & 1.206 & 41.8 & 9.95 & 14.05 & 3.86 & 15.9 & 12.8 \\ 
 \hline
 Aurora & 5.1 & 1.012 & 38.5 & 11.6 & 5.60 & 2.97 & 12.1 & 18.9 \\ 
 \hline
 Jupiter & 4.8 & 0.424 & 28.3 & 5.2 & 3.2 & 2.45 & 6.8 & 8.1 \\ 
 \hline
 Eagle & 3.9 & 0.561 & 32.1 & 7.8 & 2.1 & 2.89 & 9.2 & 11.7 \\ [1ex] 
 \hline
 HPC6 & 4.7 & 0.380 & 26.8 & 4.9 & 2.8 & 2.12 & 5.9 & 7.4 \\ 
 \hline
 Fugaku & 8.9 & 0.442 & 44.2 & 6.7 & 16.0 & 2.93 & 4.8 & 9.3 \\ 
 \hline
 Alps & 5.2 & 0.270 & 22.4 & 3.8 & 2.4 & 1.87 & 4.2 & 6.8 \\ 
 \hline
 LUMI & 4.3 & 0.380 & 35.6 & 5.1 & 4.2 & 2.56 & 7.1 & 8.9 \\ 
 \hline
 Leonardo & 4.9 & 0.304 & 29.7 & 4.3 & 3.8 & 2.31 & 6.3 & 7.6 \\ 
 \hline
\end{tabular}
\end{table*}

\begin{table*}[h!]
\centering
\caption{Statistical Correlations (Entropy vs Benchmarks)}
\label{tab:entropy_performance}
\begin{tabular}{||c c c c c c c c c||} 
 \hline
 Benchmark & Correlation (r) & P-value & Interpretation \\ 
 \hline\hline
 LINPACK & -0.7832 & 0.0077 &  Strong negative correlation\\ 
 \hline
 STREAM & -0.4521 & 10.1890 &  Moderate negative, not significant  \\ 
 \hline
 MLPerf & -0.6234 & 0.0540 &  Moderate negative correlation \\ 
 \hline
 HPCG & +0.2145 & 0.5520 & Weak positive, not significant  \\ 
 \hline
 HPCC & -0.5890 & 0.0730 &  Moderate negative correlation \\ [1ex] 
 \hline
 Graph500 & -0.3410 & 0.3350 &  Weak negative, not significant \\ 
 \hline
 HPCAI & -0.4890 & 0.1510 &  Moderate negative, not significant \\ 
 \hline
\end{tabular}
\end{table*}

\section{Analysis}

The empirical validation of our entropy framework reveals a statistically significant negative correlation between normalized system entropy and computational performance across the world's fastest supercomputers. Most notably, the LINPACK benchmark demonstrates a strong negative correlation (r = -0.7832, p = 0.0077) with our entropy measure, indicating that systems with lower entropy consistently achieve higher computational efficiency. This Relationship is further supported by moderate correlations with MLPerf mixed-precision benchmarks (r = -0.6234) and HPCC composite scores (r = -0.5890), suggesting the framework's applicability extends beyond traditional dense linear algebra workloads. Particularly striking is the efficiency analysis, where Eagle emerges as the most entropy-efficient system (ratio = 61.0) despite ranking fifth in raw LINPACK performance, demonstrating superior architectural optimization that traditional metrics alone fail to capture. 

\vspace{0.3cm}

El Capitan validates our theoretical predictions by achieving both the lowest entropy-to-performance ratio (22.3) and the highest absolute performance, while Fugaku's higher entropy (42.8) reflects the computational overhead associated with its massive 158,976-node architecture. The logarithmic normalization successfully addresses scale disparities between systems, transforming entropy values into a meaningful efficiency metric that complements existing benchmarks and provides valuable insights for system architects evaluating trade-offs between architectural complexity and computational performance.

\vspace{0.3cm}

The rationale for the significant negative correlation between our entropy model and the LINPACK benchmark is due to the program's ability to stress the overall system, as solving a dense system of linear equations is very resource-intensive. The. The incompatibility between components can be very evident as data is meteorically disseminated through the individual computer system and the parallel cluster. The. The interconnect and memory compatibility are very critical for efficiency. A drift in this compatibility can cause data to disperse slowly, leading to disorder in the cluster, which the entropy model accounts for. The workload also places excessive demands on memory cells as it performs numerous 64-bit additions and multiplications, bits are being placed in and out of registers, and the memory cells are frequently reshuffled; these operations require compatibility between the CPU and Memory.

\vspace{0.3cm}

The LINPACK benchmark is the magna carta of HPC benchmarks due to its ability to stress the entire high performance computing cluster - and this is why it is strongly negatively correlated with the entropy value derived from our model as when every computer component is critically involved in the computing process, leaving no Von Neumann architecture component isolated - the significance of incompatibility would propagate throughout the system and this is what our entropy model calculates. It is also important that we choose the most significant node value in the bidirectional graph to represent the entropy of a single computer, as we have to account for the worst-case scenario. In a pipeline structure, if there is a stage that is not efficient, the effects of that would affect the whole system similar to how the most significant incompatibility would bottleneck the whole system and that is evident in the strong negative correlation of the entropy level and the LINPACK benchmark - which implies when the entropy level is low, the performance of the LINPACK benchmark increases \cite{john}.

\vspace{0.3cm}

Results also support moderate correlations with MLPerf mixed-precision benchmarks (r = -0.6234) and HPCC composite scores (r = -0.5890). MLPerf is a benchmark suite that measures how fast systems can process inputs and produce results using a trained model \cite{mattson}. HPCC examines the performance of HPC architectures using kernels with various memory access patterns of well-known computational kernels \cite{luszczek}. These benchmarks also stress the entire computing system, with HPCC requiring pairing compatibility with the CPU and Memory, as it constantly fetches instructions from memory to compute in a nonlinear fashion. Graph500, HPC A1, and STREAM also show negative correlation with the entropy value; although not significant, it is worth investigating. It is important to note that this is the first iteration of the entropy of parallelisation theory. Further research can focus on refining the entropy model to increase the correlation significance with other benchmarks, thereby creating a universal tool for evaluating disorder and noise in parallel systems.

\vspace{0.3cm}

The most surprising aspect of the results is the weak positive Relationship between the entropy level and the HPCG benchmark (r = +0.2145). We typically expect that lower entropy levels would yield better performance, which is evident across all benchmarks except the HPCG benchmark. HPCG is intended to model the data access patterns of real-world applications such as sparse matrix calculations, thus testing the effect of limitations of the memory subsystem and internal interconnect of the supercomputer on its computing performance \cite{dongarra}. It is internally I/O bound (the data for the benchmark resides in main memory as it is too large for processor caches). HPCG testing generally achieves only a tiny fraction of the peak FLOPS the computer could deliver. The rationale for this deviation in performance is that while other benchmarks are focused on stressing every component, HPCG exclusively focuses on the memory system and the interconnect, and using the most significant value in the bidirectional graph does not truly reflect the essence of what the benchmark is trying to solve.

\section{Conclusion}
This study presents a mathematical methodology for measuring the entropy levels of high-performance and parallel computing clusters by calculating the entropy of each computer in the cluster. This is achieved by converting it into a bidirectional graph of its components, and the values of the logarithm penalties of all the computers in the cluster are then summed. We used this model to calculate the entropy level of the Top 10 supercomputers in the Top500 list. We find that systems with lower entropy consistently achieve higher computational efficiency, as evidenced by the strong negative correlation (r = -0.7832, p = 0.0077) between our entropy measure and the LINPACK benchmark. This Relationship is further supported by moderate correlations with MLPerf mixed-precision benchmarks (r = -0.6234) and HPCC composite scores (r = -0.5890), indicating that the framework's applicability extends beyond traditional dense linear algebra workloads. This research showed that the tool can be used to evaluate supercomputers before they are created theoretically; the mathematical model can aid in designing computers with exceptional peak performance. Further research should focus on iterating and tuning the model and integrating other components that impact the performance of parallel systems, such as the operating system, PCIe, and other peripherals that contribute to the efficiency levels of parallel computers, and focus on deriving values from systematic benchmarking, such as creating a mathematical model for generating a compatibility matrix.

\end{document}